\documentstyle[12pt]{article}
\begin{document}


\renewcommand{\thesection}{\arabic{section}}
\renewcommand{\theequation}{\arabic{equation}}
\renewcommand {\c}  {\'{c}}
\newcommand {\cc} {\v{c}}
\newcommand {\s}  {\v{s}}
\newcommand {\CC} {\v{C}}
\newcommand {\C}  {\'{C}}
\newcommand {\Z}  {\v{Z}}
\newcommand{\pv}[1]{{-  \hspace {-4.0mm} #1}}

\baselineskip=24pt


\begin{center}
{\bf   Solitons in the Calogero model for distinguishable particles}
 
\bigskip

V. Bardek{ \footnote{e-mail: bardek@irb.hr}} \hspace{0.3cm} and \hspace{0.3cm}
S.Meljanac {\footnote{e-mail: meljanac@irb.hr}} \\  
 Rudjer Bo\v{s}kovi\'c Institute, Bijeni\v cka  c.54, HR-10002 Zagreb,
Croatia\\[3mm]

\bigskip

\end{center}
\setcounter{page}{1}
\bigskip


\begin{center}
{\bf   Abstract}

\bigskip

\end{center} 

 We consider a large $- N, \; $ two-family Calogero model in the Hamiltonian,
collective-field approach. The Bogomol'nyi limit
appears and the corresponding solutions are given by the 
static-soliton configurations. Solitons from different families are localized at the same place.
They behave like a paired hole and lump on the top
of the uniform vacuum condensates, depending on the values of the
coupling strengths. When the number of particles in the first family is much larger than that of the second family,
 the hole solution goes to the vortex profile already found in the one-family Calogero model.

\bigskip
PACS number(s): 03.65.Sq, 05.45.Yv, 11.10.Kk, 11.15.Pg \\
\bigskip
\bigskip
Keywords: multispecies Calogero model, collective-field theory, solitonic solutions


\newpage




The Calogero system is a class of exactly solvable models in one dimension [1-3]. 
The model has found wide applications in areas as diverse as condensed
matter \cite{Haldane:1987gg}, black-hole physics \cite{Gibbons:1998fa} and two-dimensional string theory
 \cite{Gomis:2003vi}.
 The ordinary Calogero model describes $ \; N \; $
indistinguishable particles on the line which interact through an inverse-square two-body interaction.
As far as the distinguishable particles are concerned there are a few generalizations of the Calogero model to models
 of particles with different masses and with $ \; \frac{1}{ r_{ij}^{2}} \; $ couplings depending on the labelling of the 
particles coupled [7-14]. 
A multispecies one-dimensional Calogero model with two- and three-body
interactions was treated in the $ \; SU(1,1) \; $  algebraic approach in \cite{Meljanac:2003jj,Meljanac:2004vj}, while
its matrix formulation was presented in \cite{Meljanac:2004mr}. Although it was possible to
find an infinite number of exact eigenstates and eigenenergies, the set is
not complete.

 Recently, it was shown in Ref. \cite{Guhr:2004ff} that a natural 
supersymmetric extension of the Calogero model resulted in two-family Calogero models which are 
 exactly solvable in some special cases.

In an attempt to better understand the nature of the multispecies Calogero model, in this paper we transform
the two-family model to collective fields. This transformation gives much more insight into the non-perturbative,
solitonic sector of the theory.

Let us start with the Calogero Hamiltonian describing two different families of particles in
 interaction \cite{Meljanac:2004vj}:
$$
 H = - \frac{1}{2 m_{1}} \sum_{i}^{N_{1}} \frac{{\partial}^{2}}{\partial {x_{i}}^{2}} +
    \frac{\lambda (\lambda - 1)}{2 m_{1}} \sum_{i \neq j }^{N_{1}} \frac{1}{{(x_{i} - x_{j})}^{2}} -
     \frac{1}{2 m_{2}} \sum_{\alpha}^{N_{2}} \frac{{\partial}^{2}}{\partial {x_{\alpha}}^{2}} +
    \frac{\nu (\nu - 1)}{2 m_{2}} \sum_{\alpha \neq \beta }^{N_{2}} \frac{1}{{(x_{\alpha} - x_{\beta})}^{2}}  
$$
$$
 + \frac{1}{2} \sum_{i}^{N_{1}}\sum_{\alpha }^{N_{2}}
 \frac{\kappa(\kappa -1)}{(x_{i}-x_{\alpha})^{2}}\left(
\frac{1}{ m_1}  +  \frac{1} { m_2 } \right) +
$$
$$
+ \frac{1}{2}\sum_{i}^{N_{1}} \sum_{\alpha \neq \beta}^{N_{2}}
 \left( \frac{{\kappa}^{2}}
      {m_{1}(x_{i}-x_{\alpha}) (x_{i}-x_{\beta})}\right) +    
\sum_{i}^{N_{1}} \sum_{\alpha \neq \beta}^{N_{2}} \left(\frac{\nu \kappa}
      {m_{2} (x_{\alpha}-x_{i}) (x_{\alpha}-x_{\beta})} \right) 
$$
\begin{equation}
 + \frac{1}{2}\sum_{i \neq j}^{N_{1}} \sum_{\alpha}^{N_{2}}
 \left( \frac{{\kappa}^{2}}
      {m_{2} (x_{\alpha} - x_{i}) (x_{\alpha} - x_{j})} \right )+  
\sum_{i \neq j}^{N_{1}} \sum_{\alpha}^{N_{2}} \left( \frac{\lambda \kappa}
      {m_{1} (x_{i}-x_{\alpha}) (x_{i}-x_{j})} \right ). 
\end{equation}

The first family contains $ \; N_{1} \; $ particles with mass $ \; m_{1} \; $ at positions
 $ \; x_{i}, \; i = 1,...,N_{1}, \; $ while the second one contains 
$ \; N_{2} \; $ particles with mass $ \; m_{2} \; $ at positions
 $ \; x_{\alpha}, \; \alpha = 1,...,N_{2}. $ The particles of the same kind interact and the corresponding coupling
constants within each family are given by  $ \; \lambda \; $ and $ \; \nu, \; $ respectively. The particles of 
different kind also interact and the interaction strength between the first and the second family is denoted by
$ \; \kappa. $ We consider the parameters $ \; \lambda, \nu \; $ and $ \; \kappa \; $ positive.

The Hamiltonian (1) describes the simplest multispecies Calogero model for particles on the line, interacting with the 
two-  and three-body potentials. Setting $ \; \lambda = \nu = \kappa \; $ and $ m_{1} = m_{2}, \; $ we recover the
ordinary $ \; N- $ body Calogero model. The three-body terms in (1) trivially vanish in this case.
General conditions for the absence of three-body interactions are given in \cite{Meljanac:2003jj,Meljanac:2004vj}.
 In the following we do not use any confining potentials.

 We can perform the similarity transformation
\begin{equation}
 H \rightarrow \Pi_{\kappa}^{-1} \Pi_{\nu}^{-1} \Pi_{\lambda}^{-1} H \Pi_{\lambda} \Pi_{\nu} \Pi_{\kappa}
\end{equation}
to obtain a simpler but non-hermitian Hamiltonian
$$
 H = - \frac{1}{2 m_{1}} \sum_{i}^{N_{1}} \frac{{\partial}^{2}}{\partial {x_{i}}^{2}} -
    \frac{1}{m_{1}} \left( \lambda \sum_{i \neq j}^{N_{1}}
   \frac{1}{x_{i} - x_{j}} + \kappa \sum_{i, \alpha} \frac{1}{x_{i} - x_{\alpha}} \right) \frac{\partial}{\partial x_{i}}
$$
\begin{equation}
    - \frac{1}{2 m_{2}} \sum_{\alpha}^{N_{2}} \frac{{\partial}^{2}}{\partial {x_{\alpha}}^{2}} -
    \frac{1}{m_{2}} \left( \nu \sum_{\alpha \neq \beta}^{N_{2}}
   \frac{1}{x_{\alpha} - x_{\beta}} + \kappa \sum_{i, \alpha} \frac{1}{x_{\alpha} - x_{i}} \right)
        \frac{\partial}{\partial x_{\alpha}},
\end{equation}
where the two-  and three-body interactions have simply disappeared.
The Jastrow prefactors are given by
$$
 \Pi_{\lambda} = \prod_{i < j}^{N_{1}} { (x_{i} - x_{j})}^{\lambda},
$$
$$
 \Pi_{\nu} = \prod_{\alpha < \beta}^{N_{2}} {( x_{\alpha} - x_{\beta})}^{\nu},
$$
\begin{equation}
 \Pi_{\kappa} = \prod_{i, \alpha}^{N_{1}N_{2}} {( x_{i} - x_{\alpha})}^{\kappa}
\end{equation}
and incorporate the conditions that the wave functions go to zero whenever the particles approach each other.
 The usual approach to the quantum-mechanical problem is to solve the eigenvalue problem with the Hamiltonian (1) using 
symmetric and  antisymmetric wave functions, depending on the underlying statistics of the identical particles.
 Instead, we 
develop a collective-field theory of this system in the large $- N_{1} \; $ and large $- N_{2} \; $ sectors of the 
Hilbert space.

The collective-field theory for the two-family Calogero model is obtained by changing variables from the
particle coordinates $ \; x_{i} \; $ and $ \; x_{\alpha} \; $  to the density fields $ \; \rho(x) \; $ and
$ \; \tilde{\rho}(x) \; $  defined as
\begin{equation}
 \rho(x) = \sum_{i = 1}^{N_{1}} \delta( x - x_{i}),
\end{equation}
\begin{equation}
 \tilde{\rho}(x) = \sum_{\alpha = 1}^{N_{2}} \delta( x - x_{\alpha}).
\end{equation}
Such a change of variables is meaningful only if the particle numbers $ \; N_{1} \; $ and $ \; N_{2} \; $ go to
infinity \cite{Jevicki:1979mb,Andric:1991wp,Andric:1994nc}.
The Hamiltonian (3) can be expressed entirely in terms of $ \; \rho(x), \; \tilde{\rho}(x) \; $ 
and their canonical conjugates
\begin{equation}
  \pi(x) = - i \frac{\delta}{\delta \rho(x)},
\end{equation}
\begin{equation}
 \tilde{ \pi}(x) = - i \frac{\delta}{\delta \tilde{\rho}(x)},
\end{equation}
satisfying the following equal-time commutation relations:
\begin{equation}
 [ \rho(x), \pi(y)] = i \delta(x - y),
\end{equation}
\begin{equation}
 [\tilde{ \rho}(x),  \tilde{\pi}(y)] = i \delta(x - y),
\end{equation}
\begin{equation}
 [ \rho(x), \tilde{\rho}(y)] = [ \pi(x), \tilde{\pi}(y)] = 0.
\end{equation}
After the change of the variables, the Hamiltonian (3) takes the form
$$
 H = \frac{1}{2 m_{1}} \int dx \rho(x) { ( \partial_{x} \pi(x))}^{2}
$$
$$ 
 - \frac{i}{m_{1}} \int dx \rho(x) \left( \frac{\lambda - 1}{2} \frac{\partial_{x} \rho}{\rho} +
 \lambda \pv \int \frac{ dy \rho(y)}{x - y}  + \kappa \pv \int \frac{dy \tilde{\rho}(y)}{x - y} \right) \partial_{x} \pi(x)
$$
$$
 + \frac{1}{2 m_{2}} \int dx \tilde{\rho}(x){(\partial_{x} \tilde{\pi}(x))}^{2} 
$$
\begin{equation}
 - \frac{i}{m_{2}} \int dx \tilde{ \rho}(x) \left( \frac{\nu - 1}{2} \frac{\partial_{x} \tilde{ \rho}}{\tilde{\rho}}
  + \nu \pv \int \frac{ dy \tilde{ \rho}(y)}{x - y}
  + \kappa \pv \int \frac{dy \rho(y)}{x - y} \right) \partial_{x} \tilde{ \pi}(x),
\end{equation}
where $ \; \pv \int \; $ denotes Cauchy's principal value of the integral. This Hamiltonian is still non-hermitian
owing to the imaginary terms. In order to obtain the hermitian Hamiltonian, we have to rescale  Schrodinger's wave 
functions of the original Hamiltonian by using the Jacobian of the transformation from $ \; \{ x_{i}, x_{\alpha} \} \; $
to $ \; \{ \rho(x), \tilde{\rho}(x) \}, \;  $ as was suggested in Ref. \cite{Jevicki:1979mb}.
After performing a straightforward algebra, we find the Jacobian $ \; J $ 
$$
 ln J = (1 - \lambda) \int dx \rho(x) ln \rho(x) + (1 - \nu) \int dx \tilde{\rho}(x) ln \tilde{\rho}(x)
$$
$$
 - \lambda \int dx dy \rho(x) ln |x - y | \rho(y) - \nu \int dx dy \tilde{ \rho}(x) ln |x - y | \tilde{ \rho}(y)
$$
\begin{equation}
  - 2\kappa \int dx dy \rho(x) ln |x - y | \tilde{ \rho}(y).
\end{equation}
The hermitian Hamiltonian is finally given by
\newpage
$$
 H \rightarrow J^{\frac{1}{2}} H J^{- \frac{1}{2}} 
$$
$$
 = \frac{1}{2 m_{1}} \int dx \rho(x) { ( \partial_{x} \pi(x))}^{2} +
  \frac{1}{2 m_{1}} \int dx \rho(x) {\left( \frac{\lambda - 1}{2} \frac{\partial_{x} \rho}{\rho} +
 \lambda \pv \int \frac{ dy \rho(y)}{x - y}  + \kappa \pv \int \frac{dy \tilde{\rho}(y)}{x - y} \right)}^{2}
$$
$$
 + \frac{1}{2 m_{2}} \int dx \tilde{\rho}(x){(\partial_{x} \tilde{\pi}(x))}^{2} 
 + \frac{1}{2 m_{2}} \int dx \tilde{ \rho}(x) {\left( \frac{\nu - 1}{2} \frac{\partial_{x} \tilde{ \rho}}{\tilde{\rho}}
  + \nu \pv \int \frac{ dy \tilde{ \rho}(y)}{x - y}
  + \kappa \pv \int \frac{dy \rho(y)}{x - y} \right)}^{2} -
$$
$$
  - \frac{\lambda}{2 m_{1}}  \int dx \rho(x) \partial_{x} \left. \frac{P}{x - y} \right|_{y = x}  -
 \frac{\lambda - 1}{4 m_{1}} \int dx {\partial_{x}}^{2} \left. \delta(x - y) \right|_{y = x} -
$$
\begin{equation}
 - \frac{\nu}{2 m_{2}} \int dx \tilde{\rho}(x) \partial_{x} \left. \frac{P}{x - y} \right|_{y = x}
- \frac{\nu - 1}{4 m_{2}} \int dx {\partial_{x}}^{2} \left. \delta(x - y) \right|_{y = x},
\end{equation}
where $ \; P \; $ stands for the principal part.
The two terms, quadratic in the conjugate momentum operators $ \; \pi \; $ and $ \; \tilde{\pi}, \; $ represent
the kinetic energy of the system. The rest emerges as a quantum collective-field potential. The last terms are 
singular and do not give a contribution in the leading order in the $ \; \frac{1}{N_{1}}  \; $ and
$ \; \frac{1}{N_{2}}  \; $ expansions. They should be cancelled by the infinite zero-point energy of the collective
fields $ \; \rho \; $ and $ \; \tilde{ \rho}. $

To find the ground-state energy of our system, we assume that the corresponding densities are static. Since their
 momenta are vanishing, the leading part of the Hamiltonian in the $ \; \frac{1}{N_{1}}  \; $ and
$ \; \frac{1}{N_{2}}  \; $ expansions is given by the effective potential
$$ 
 V_{eff}(\rho, \tilde{\rho})  = 
 \frac{1}{2 m_{1}} \int dx \rho(x) {\left( \frac{\lambda - 1}{2} \frac{\partial_{x} \rho}{\rho} + 
 \lambda \pv \int \frac{ dy \rho(y)}{x - y}  + \kappa \pv \int \frac{dy \tilde{\rho}(y)}{x - y} \right)}^{2} 
$$
\begin{equation}
 +  \frac{1}{2 m_{2}} \int dx \tilde{ \rho}(x) {\left( \frac{\nu - 1}{2} \frac{\partial_{x} \tilde{ \rho}}{\tilde{\rho}}
  + \nu \pv \int \frac{ dy \tilde{ \rho}(y)}{x - y}
  + \kappa \pv \int \frac{dy \rho(y)}{x - y} \right)}^{2}.
\end{equation}
Its form makes the Bogomol'nyi bound apparent. 
The potential is positive semi-definite and its contribution to the ground-state energy vanishes if there exist positive
solutions of the coupled equations
$$
  \frac{\lambda - 1}{2} \frac{\partial_{x} \rho}{\rho} +
 \lambda \pv \int \frac{ dy \rho(y)}{x - y}  + \kappa \pv \int \frac{dy \tilde{\rho}(y)}{x - y} = 0, 
$$
\begin{equation}
\frac{\nu - 1}{2} \frac{\partial_{x} \tilde{ \rho}}{\tilde{\rho}}
  + \nu \pv \int \frac{ dy \tilde{ \rho}(y)}{x - y}
  + \kappa \pv \int \frac{dy \rho(y)}{x - y} = 0.
\end{equation}
It is evident that there always exist uniform solutions
\begin{equation}
  \rho(x) = \rho_{0}, \;\;\;\;\; \tilde{\rho}(x) = \tilde{\rho}_{0}.
\end{equation}

We have not been able to obtain analytic solutions to these equations for any values of the parameters $ \; \lambda,
\nu \; $ and $ \; \kappa. $ However, if we further simplify our model by the assumption that there are no 
three-body interactions between the particles in the starting Hamiltonian (1),
we obtain the conditions ( see Ref. \cite{Meljanac:2004vj} )
$$
  {\kappa}^{2} = \lambda \nu,
$$
\begin{equation}  
   {\left( \frac{m_{2}}{m_{1}} \right) }^{2} = \frac{\nu}{\lambda}.
\end{equation}

In this particular case, the solutions of equations (16) are always interrelated by
\begin{equation}
 {\tilde{\rho}}^{{\nu}^{\frac{1}{2}} - {\nu}^{- \frac{1}{2}}} \sim 
 {\rho}^{{\lambda}^{\frac{1}{2}} - {\lambda}^{- \frac{1}{2}}}. 
\end{equation}
Note that for $ \;  \lambda = \nu, \; $ the condition (19) implies proportionality
between  $ \; \rho \; $ and  $ \; \tilde{\rho} \; $  and this means that we are dealing with the one-family
model.

If we further assume that $ \; \kappa = 1 \; $ ( weak-strong coupling duality in Ref. [19] ), we end up with the condition
\begin{equation}
 \rho \tilde{\rho} = c,
\end{equation}
where $ \; c \; $ is some positive constant. This very condition allows us to find new
 soliton solutions to the coupled equations (16). In fact, there is only one relevant equation, let us say for 
$ \; \rho, \; $ for example
\begin{equation}
 \frac{\lambda - 1}{2} \frac{\partial_{x} \rho}{\rho} +
 \lambda \pv \int \frac{ dy \rho(y)}{x - y}  + c \pv \int \frac{dy }{ \rho(y) (x - y)} = 0.
\end{equation}
This equation can be solved by a rational  {\em{Ansatz}}
\begin{equation}
  \rho(x) = \rho_{0} \frac { x^{2} + a^{2}}{x^{2} + b^{2}},
\end{equation}
where $ \; a \; $ and $ \; b \; $ are positive constants. By using the Hilbert transform 
\begin{equation}
 \pv \int \frac {dy}{x - y} \frac{1}{y^{2} + a^{2}} = \frac{\pi}{a} \frac{x}{x^{2} + a^{2}},
\end{equation}
we find the conditions
$$
  \lambda - 1 +  \frac{c \pi}{\rho_{0} a} ( b^{2} - a^{2}) = 0, 
$$
\begin{equation}
 1 - \lambda + \frac{\lambda \rho_{0} \pi }{b} ( a^{2} - b^{2}) = 0. 
\end{equation}
The soliton solution for the first family is given by
\begin{equation}
  \rho(x) = \rho_{0} + \frac{\lambda - 1}{\lambda \pi} \frac{b}{x^{2} + b^{2}},
\end{equation}
while the solution for the second family looks like
\begin{equation}
 \tilde{\rho}(x) = \frac{c}{\rho(x)} = \frac{c}{ \rho_{0}} + \frac{1 - \lambda}{ \pi} \frac{a}{x^{2} + a^{2}}.
\end{equation}
We note that both solutions are localized at $ \; x = 0. $
For large values of $ \; x, \; $  the soliton solutions approach the uniform solutions found before.
This yields one more condition
\begin{equation}
 \frac{\rho (\infty)}{\tilde{\rho} (\infty)} = \frac{N_{1}}{N_{2}} = \frac{ {\rho_{0}}^{2}}{c},
\end{equation}
which, together with the conditions (24), finally fixes the parameters $ \; a, b \; $ and $ \; c,$ namely
$$
 a = \frac {N_{1} ( \lambda -1)}{ \rho_{0} \pi N_{2} ( 1 -  {N_{1}}^{2} { \lambda}^{2} / {N_{2}}^{2} )},
$$
$$
 b = \frac { (1 -  \lambda)}{\lambda \rho_{0} \pi  ( 1 -  {N_{2}}^{2} /  { \lambda}^{2}{N_{1}}^{2} )},
$$
\begin{equation}
 c = {\rho_{0}}^{2} \frac{N_{2}}{N_{1}}.
\end{equation}

The particle number of the first soliton is 
\begin{equation}
  \int dx ( \rho(x) - \rho_{0} ) = \frac{\lambda - 1}{\lambda},
\end{equation}
while that of the second soliton is
\begin{equation}
  \int dx ( \tilde{\rho}(x) - \tilde{\rho}_{0} ) = 1 - \lambda,
\end{equation}
where $ \; \tilde{\rho}_{0} = \frac{c}{\rho_{0}}. $ It is interesting to observe that these numbers are generally not
 integers.

For $ \; \lambda < 1, \; $ the first soliton behaves like the hole in the condensate $ \; \rho_{0} \; $ and the second one 
behaves like the particle above the condensate $ \; \tilde{\rho}_{0}. $ The roles are interchanged for $ \; \lambda > 1. $
It is interesting to study the limiting case $ \; c \rightarrow 0. $
This can be achieved only if $ \; N_{1} \; $ is much larger than $ \; N_{2}.$ The condition (24) gives 
$ \; a \rightarrow 0 \; (\mbox{the case} \; b \rightarrow \infty \; \mbox{ destroys the solution}).  $ The first-family
soliton reduces to the ''vortex'' profile
\begin{equation}
  \rho(x) = \rho_{0} \frac { x^{2}}{x^{2} + b^{2}}, \quad  b = \frac{1 - \lambda}{\lambda \pi \rho_{0}}.
\end{equation}
The above solution is called a vortex because the density goes to zero at $ \; x = 0. $
It has already been obtained in the collective-field approach to the
one-family Calogero model \cite{Andric:1994nc,Polychronakos:1994xg}.
The second-family lump solution transforms into the sharp delta-function profile
\begin{equation}
 \tilde{\rho}(x) = (1 - \lambda) \delta(x).
\end{equation}
In deriving this result we have used the Lorentzian representation of a delta function
\begin{equation}
 \delta(x) = \lim_{\varepsilon \rightarrow \infty} \frac{1}{\pi} \frac{\varepsilon}{x^{2} + {\varepsilon}^{2}}.
\end{equation}

We have thus demonstrated that our solitons have vanishing energy in the leading approximation. It does not cost
 any energy to create such a pair of correlated solitons.

What about the other solutions to the coupled equations (16)? The authors of Ref. \cite{Andric:2002sp}
 have claimed that there exists
a multi-vortex solution in the first family accompanied by the sum of the  delta-function profiles for the second
family. The zeros of the $ \; \rho(x) \; $ (i.e. the positions of the vortices) are simultaneously the points
at which $ \; \tilde{ \rho}(x) \; $ diverges (i.e. the positions of the particles).
According to Ref. \cite{Andric:2002sp}, ''the collective field $ \; \tilde{\rho}(x) \; $ describes solitons as
particles, whereas $ \; \rho(x) \; $ gives a microscopic description in terms of fields''.
 This is the essence of the so-called
particle-vortex duality.
We do not agree with this interpretation since $ \; \rho(x) \; $  and  $ \; \tilde{\rho}(x) \; $
describe two different physical systems and, consequently, cannot be
applied simultaneously to the same object ( soliton in this particular
case). Moreover, we doubt that there exist multi-vortex solutions at least not of
the form suggested in Ref. \cite{Andric:2002sp}!
Nevertheless, let us investigate this possibility more carefully in the case of the hypothetical two-vortex 
solution. This solution can be reached by the limiting procedure $ \; (c \rightarrow 0 ), \; $ starting from the 
two-hole solution in the condensate $ \; \rho_{0} $
\begin{equation}
 \rho(x) = \rho_{0} \frac{ (x^{2} - a^{2}) (x^{2} - \bar{a}^{2})} { (x^{2} - b^{2}) (x^{2} - {\bar{b}}^{2})},
\end{equation}
where $ \; a, \bar{a} \; $ and $ \; b, \bar{b} \; $ are complex parameters.
This {\em{Ansatz}}$ \; $  for $ \; \rho(x) \; $ takes into account the
 positiveness of the particle density automatically.
Furthermore, the rational function of two polynomials with the fourth degree in $ \; x \; $ in principle ensures
the two-hole ( and for $ \; \tilde{\rho} = \frac{c}{\rho}, \; $ the two-lump) structure.

Inserting the expression (34) for $ \; \rho \; $ and $ \;  \frac{c}{\rho} \; $ for $ \; \tilde{\rho} \; $ in equations
(16), and by using the Hilbert transform
\begin{equation}
  \pv \int \frac {dy}{x - y} \frac{1}{y^{2} - a^{2}} = \pm  \frac{i \pi x}{a (x^{2} - a^{2})},
\end{equation}
where the $ \; + \; $ sign ( $- \; $ sign) is taken for $ \; a \; $ in the upper (lower) half of the complex plane, we
obtain the following system of algebraic equations:
$$ 
 \lambda - 1 \pm \frac{c \pi i}{\rho_{0} a }( a^{2} - b^{2}) \frac{ a^{2} - \bar{b}^{2}} { a^{2} - \bar{a}^{2}} = 0,
$$
\begin{equation}
 1 - \lambda  \pm \frac{\lambda \rho_{0} \pi i}{b }( b^{2} - a^{2}) \frac{ b^{2} - \bar{a}^{2}} { b^{2} - \bar{b}^{2}} = 0.
\end{equation}

Let us set 
\begin{equation}
 a = \alpha + i \beta, \;\;\;\;\; b = \gamma + i \delta,
\end{equation} 
where $ \; \alpha, \beta, \gamma \; $ and $ \; \delta \;$ are real numbers. The real and the imaginary parts of the 
complex equations (36) are, respectively,
$$
  \lambda - 1 \pm \frac{c \pi}{\rho_{0} \beta} ( \alpha^{2} - \beta^{2} - \gamma^{2} + \delta^{2}) = 0,
$$
$$
 1 - \lambda \pm \frac{\lambda \pi \rho_{0}}{\delta} (\gamma^{2} - \delta^{2} - \alpha^{2} + \beta^{2}) = 0,
$$
$$
 4 \alpha^{2} \beta^{2} - 4 \gamma^{2}\delta^{2} + ( \alpha^{2} -  \beta^{2} - \gamma^{2} + \delta^{2})
   (3 \alpha^{2} +  \beta^{2} + \gamma^{2} - \delta^{2}) = 0,
$$   
\begin{equation}
 4 \gamma^{2}\delta^{2} - 4 \alpha^{2} \beta^{2} + (  \gamma^{2} - \delta^{2} - \alpha^{2} +  \beta^{2})
                           ( 3 \gamma^{2} + \delta^{2} + \alpha^{2} -  \beta^{2}) = 0.
\end{equation}
The only solutions are 
$$
 \alpha = \gamma = 0, \;\;\;\;  \beta^{2} = \delta^{2},
$$
$$
\alpha^{2} = \gamma^{2}, \;\;\;\;   \beta = \delta = 0,
$$
\begin{equation}
 \alpha = \beta = \gamma = \delta = 0.
\end{equation}
This brings us back to the uniform solution $ \; \rho(x) = \rho_{0}. $ Similarly, one can show that the {\em{Ansatz}}
$ \;  \rho(x) = \rho_{0} \frac{ (x^{2} + a_{1}^{2}) (x^{2} + a_{2}^{2})}
       { (x^{2} + b_{1}^{2}) (x^{2} + b_{2}^{2})}, \; $ for real parameters $ \; a_{1}, b_{1}, a_{2} \; $ and
$ \; b_{2}, \; $ leads to the one hole-lump pair solution, Eqs. (25) and (26).

In conclusion, the only duality that can be considered is the trivial
self-duality of the collective Hamiltonian (14). Obviously, it is invariant
under the interchange of the two families, namely $ \; \rho \leftrightarrow \tilde{\rho}, \;\;
 \pi \leftrightarrow \tilde{\pi}, \;\;  \lambda \leftrightarrow \nu  \;\;  \mbox{ and} \;\;  m_{1} \leftrightarrow m_{2}. $

Our final remark is that our two-family Calogero model without the three-body interaction (18) can be viewed
 as the one-family model, but only in the leading approximation. Namely, comparing
 the effective potential (15) with the effective potential for the one-family Calogero model, one finds that
 they can be identified. To this end, we can define the new, effective one-family
collective field
 $ \; \rho_{eff} =  {( \frac{m}{{\mu}^{2} \lambda m_{1}})}^{1/3} ( \lambda \rho + \kappa \tilde{\rho} ), \; $
 describing one-family
particles with the effective mass $ \; m \; $ and effective one-family coupling strength  $ \; \mu. $ 
Note that this identification cannot fix the values of the afore-mentioned parameters.
 However, owing to the presence of the $ \; \partial_{x} \rho / \rho \; $ and
 $ \; \partial_{x} \tilde{ \rho} / \tilde{ \rho} \; $ terms,
which are suppressed, respectively, by the factors $ \; \frac{1}{N_{1}} \; $ and $ \; \frac{1}{N_{2}} \; $
 with respect to the leading terms,
this equivalence breaks down. The derivative terms are crucial for the appearance of the soliton solutions.
 In this respect,
the two-family models display the new, paired solitons which do not exist in the one-family Calogero models.

Although we have here focused on a two-family Calogero model, the discussion
in this paper should be relevant to two- matrix models, since it is
known that the Calogero model actually corresponds to the $ \; O(N), \;  U(N) \; $ and
 $ \; Sp(N) \; $ invariant matrix models for $ \; \lambda = \frac{1}{2},1,2, \; $ respectively [21].
Our results can also be easily extended to the models with more than two
 distinct families of distinguishable particles. The open problems
that still remain are the question of the quantum stability of the
semi-classical solutions (25) and (26), and the existence of the possible
moving-soliton solutions. We hope to study these issues in the near future.

\bigskip
\bigskip



{\bf Acknowledgment}\\
This work was supported by the Ministry of Science and Technology of the Republic of Croatia under 
contract No. 0098003.

\newpage




\end{document}